\documentstyle[preprint,aps,eqsecnum,psfig]{revtex}
\begin{document}

\draft
{\tighten
\preprint{\vbox{\hbox{U. of Iowa preprint 002501}}}

\title{Non-Gaussian numerical errors versus mass hierarchy}

\author{Y. Meurice, and M.B. Oktay}

\address{Department of Physics and Astronomy\\
The University of Iowa, Iowa City, Iowa 52242, USA}

\maketitle

\begin{abstract}

We probe the numerical errors made in renormalization group 
calculations by varying slightly the rescaling 
factor of the fields and rescaling back in order to get the 
same (if there were no round-off errors) zero momentum 2-point function
(magnetic susceptibility).
The actual calculations were performed with Dyson's hierarchical 
model and a simplified version of it.
We compare the 
distributions of numerical values
obtained from a large 
sample of rescaling factors with the (Gaussian by design) distribution 
of a 
random number generator and find significant departures from the Gaussian
behavior. In addition, the average value differ
(robustly) from the exact answer by a quantity which is of the 
same order as the standard deviation.
We provide a simple model 
in which the errors made at shorter distance have a larger weight 
than those made at larger distance. 
This model explains in part 
the non-Gaussian features and 
why the central-limit theorem does not apply.

\end{abstract}

\newpage
\section{Introduction}

In the standard model of electro-weak and strong interactions, all
the masses are generated by couplings to a scalar Higgs particle. 
However,
one might be inclined to think that, ultimately, gravitational interactions
should be responsible for mass generation. If there is no new physics 
between the electro-weak scale and the Planck scale and if we know 
nothing about the Planck scale interactions, the best we can do is
to use an effective scalar Lagrangian with a cut-off of the order 
of the Planck 
scale. In this approach, one is confronted with the hierarchy
problem\cite{susskind}, 
which means that in order to keep a physical mass very small
in cut-off units, one needs to fine-tune some parameter of the 
theory (usually the bare mass) with an incredible precision. 
In four dimensions, this fine-tuning
can be understood in terms of 
perturbative quadratic divergences. 

In the renormalization 
group (RG) approach of field theory\cite{wilson}, 
the need for fine-tuning can be related to the existence of 
unstable directions in a way which is 
independent of perturbation theory in a particular dimension. 
In this approach, the renormalized mass expressed 
in cut-off units decreases 
exponentially with the number of iterations spent near an unstable 
fixed point. In order to keep the mass small, one needs to fine-tune
the initial parameter in order to  
start close to the critical hypersurface. This hypersurface in the 
space of bare Lagrangians separates the symmetric phase from the 
symmetry broken phase and contains the unstable fixed point.

In the following, we use the statistical mechanics language where 
criticality is approached by tuning the inverse temperature $\beta$ close to 
its critical value $\beta_c$. 
In terms of this parameter, the ratio of the physical mass $m$ and the 
ultra-violet (UV)
cutoff $\Lambda$ is 
\begin{equation}
m/\Lambda\sim(\beta_c-\beta)^{\gamma/2}\ ,
\label{eq:rat}
\end{equation}
where $\gamma$ is the critical exponent associated with the 
magnetic susceptibility. In four dimensions, $\gamma=1$. If we take 
$m=100\  GeV$, a typical electroweak scale, 
and $\Lambda=10^{19}$GeV of the order of the Planck mass, 
we need to fine-tune $\beta$
with 34 digits.

The main virtue of the RG approach is to 
introduce some hierarchy in the information contained in the partition
function. At each iteration, the information relevant to understand the 
large distance behavior is amplified, while the rest of the information
is discarded according to its degree of irrelevance. 
However if some ``noise'' is introduced in this process, for instance 
as numerical errors in the calculation, the error in the relevant 
direction will be amplified too. This may lead to situations 
where the amplified errors wipe out the 
careful fine-tuning and one obtains meaningless results. 

In this article, we discuss the numerical errors
with examples where the RG transformation can be calculated 
for many iterations. 
The models used for our calculations are Dyson's hierarchical
model \cite{dyson}, 
for which very accurate methods of calculation \cite{finite,gam3}
have been recently designed,
and a simplified version \cite{toy} 
of this model for which the  
nonlinear aspects of the interpolation between fixed points is 
understood in great detail. 
For comparison, we also considered a 
random number generator designed to produce a Gaussian 
distribution.

The question of numerical errors in RG calculation has 
been briefly discussed in recent publications.
The mechanism of error amplification was identified in 
section IV of Ref. \cite{finite}. In these calculations, we  
probed  the numerical errors 
by slightly varying the parameter (denoted $s$ hereafter) 
used to rescale the 
fields at each iteration of the RG transformation 
and exploiting the fact that the physical
quantities such as the 
magnetic susceptibilty $\chi$ are independent of the choice of 
rescaling. This procedure is reviewed in section \ref{sec:strat}.
We predicted that if $\delta$ represents a typical 
round-off error (of the order of $10^{-16}$ in double precision 
calculations), 
the relative difference between the 
numerical values obtained for
two slightly different values of $s$ should be of the order
\begin{equation}
\label{eq:reler}
\Delta\chi/\chi\sim
\delta(\beta_c-\beta)^{-1}\ .
\end{equation}
This behavior was observed 
in good approximation for 
a wide range of $\beta$ (see Fig. 4 of Ref. \cite{finite}).
A consequence of this result is that at fixed 
$\delta$ and $m$, there is a maximum UV cutoff 
of the order of $m\delta^{-\gamma/2}$
beyond which the result of a calculation becomes totally
meaningless.

In the models we have considered, the exact rescaling factor $s$ for
which  the RG transformation has a non-trivial fixed point 
(denoted $s_{exact}$ hereafter) is
{\it exactly} the same as for a free theory. In models with nearest
neighbor interactions, this is only approximately the case. The 
free value is corrected 
whenever the critical exponent 
$\eta$ is non-zero. Since in most cases, $\eta$ is only known 
with a finite accuracy,
one is naturally led to consider a range of values for $s$. 
In  section III.b of 
Ref. \cite{gam3} we considered the distribution of values of the
susceptibility 
for 2000 values of $s$. 
To our surprise, we found that the errors did not 
average out to the correct value which was calculated using 
higher precision arithmetic. In addition, we found that
the distribution of errors was not Gaussian. 
In the present paper, we 
readdress these questions with larger statistics and 
we compare the results with other models. 
The technical details of our analysis 
are outlined in section \ref{sec:outline}.

Our first result is that the average 
error is of the same order of magnitude as the standard deviation 
of the distribution. There is a systematic bias in the calculations and one
does not gain anything by increasing the statistics. This bias depends on the 
peculiarities of arithmetic round-offs and we have not attempted 
to model it from a ``microscopic'' point
of view. It should however be noticed that the average error 
is ``compiler robust'' and that it should in principle be understandable 
at least for the simplified model for which only a few arithmetic operations
are needed at each iteration of the RG transformation.

Another robust result is that all our RG 
calculations  show significant 
departures from Gaussian behavior.
These departures 
can be estimated in terms of the skewness and kurtosis 
coefficients. 
In all the RG calculations 
these coefficients are at least
one order of magnitude larger than the corresponding quantities for
the Gaussian random number generator. 

How do the repeated errors 
fail to erase the details of the individual distribution 
and produce a Gaussian distribution as in the 
central limit theorem? We have considered two possible explanations.
The first one is that the distribution of errors changes from step
to step. A detailed study shows that there are indeed
small variations in these distributions,
however this might not be the most important effect.
The main reason why the central limit does not apply 
seem to be that the errors are added with 
unequal weight. The errors made during the initial iterations
are more amplified than errors made at later stage. 
As a consequence, the 
``weighted average'' inherits the skewness and kurtosis of the 
individual distribution. Since these have ``short tails'', so 
does the 
distribution of errors, unlike non-Gaussian distributions found
in the study of turbulence which have ``long tails''.

Our results impose limitations on numerical approaches of scalar field 
theory, however these are not drastic. 
The main result is that as far as the average of low 
powers of the total field is concerned, 
a hierarchy $\Lambda/m$ requires a
number of significant digits proportional to Log($\Lambda/m$), 
which is not prohibitive for $\Lambda/m\sim 10^{17}$.

\section{Technical outline of the paper}
\label{sec:outline}

The renormalization 
group (RG) approach of scalar field theory, 
the hierarchy problem and our strategy to probe the 
numerical errors are reviewed 
in generic terms in section \ref{sec:strat}.
The specific models used for our calculations are presented in 
section \ref{sec:models}. 

In section \ref{sec:distr},
we consider 
seven samples with $10^5$ data points, six from RG calculations and
one designed to produce a Gaussian distribution. It should be noted
that in all cases randomness is due to sensitive dependence on the 
initial conditions. We show that the departure from
Gaussian behavior of the distributions of RG values are significantly
larger than in the case of the random number generator 
designed to produce a Gaussian distribution.
We also show that the errors spread differently when $s<s_{exact}$
and $s>s_{exact}$, indicating that the sample should be ``resolved''
into subsamples with different distributions. 

Subsamples are analyzed quantitatively in section \ref{sec:moma}
in terms of ``uniformity indicators'' designed to establish
correlations between the values of $s$ and the moments of the 
distribution. These indicators take values close to 1 when the sample
is obtained from independent and identically distributed 
random variables.
It is found that samples with only $s<s_{exact}$
or only $s>s_{exact}$ have much better uniformity than the
the samples combining both sets of values. In the case of
the simplified model, the samples with only $s<s_{exact}$
or only $s>s_{exact}$ have values of 
the indicators which seem consistent with being samples coming
from independent and identically distributed 
random variables. For the hierarchical model, the analogous
samples need further resolution. 

It should be noted that in general one expects fluctuations of
the order of 
(spread of the parent distribution)/((size of the sample)$^{1/2}$).
Since we use large samples, these fluctuations are expected to be 
small. Compared to this small scale, the variations of the estimators
for $s<s_{exact}$
and $s>s_{exact}$
are large.
However, they are
still 
relatively small when compared to the spread itself. Consequently, 
it still makes sense to talk in an approximate way about, for instance, 
the mean of the distribution. To take an example, one can look at Table I
and see that the average errors for $s<s_{exact}$ (803)
and $s>s_{exact}$ (1297) differ by approximately one quart of the standard 
deviation of the any of the two samples. Consequently, we can still say
that for any value of $s$, the error is approximately 1000.
These remark should be kept in mind while we use expressions 
such as ``the moments of a distribution''.

A specific model where 
the errors made during the initial iterations
are more amplified than errors made at later stage
is provided in section \ref{sec:moder}.
In the conclusions, we discuss the limitations
that our results impose on numerical approaches of scalar field 
theory. The situation is  compared to what happens in chaotic
dynamical systems and in the study of turbulence.

\section{General strategy to probe the numerical errors}
\label{sec:strat}

Before getting into the specifics of model calculations,
we would like to describe in general terms, our method to 
probe the numerical errors. The main points of this section
are the following. First, the RG transformation involves a rescaling of the 
sum of the fields in boxes of increasing sizes. In general, this rescaling
factor is only known with a finite accuracy and so some range of values 
should 
in general 
be considered. Second, the physical quantities are independent
of the choice of this rescaling factor. However in practical numerical 
calculations, the arithmetic errors are different for different choices of 
rescaling factor. Consequently, we can use large samples of rescaling factors
to obtain a statistical distribution of these errors. We now proceed 
to discuss these points in the case of a generic scalar field theory.

Let us consider a scalar model 
in $D$-dimensions with a lattice spacing $a_0$. We first
integrate the fields in blocks of side $ba_0$ while keeping
the sum of the fileds in the block constant.
We then divide the sums of the fields by a factor 
$b^{(2+D-\eta)/2}$ and treat them as our new field variables. 
The procedure defines a discrete renormalization group 
transformation provided that 
the scale factor $b$ is real number strictly larger than 1.

The critical hypersurface (in the space of bare
Lagrangians) is given as the stable manifold (e.g. the basin of 
attraction) of a 
non-trivial fixed point of this
transformation. The stable manifold can be reached by considering 
a family of models indexed by a parameter which can be tuned in order
to cross the stable manifold. In field theory context, one usually pick 
the bare mass to accomplish this purpose. In the statistical mechanics 
formulation, the inverse temperature $\beta$ can be tuned to its 
critical value $\beta_c$ which is a function of the other interactions. 
This notation will be used later.

The information that we are keeping during the renormalization 
group transformation is encoded in the average values of all
the integer powers of the sum of the fields in the blocks. We call these
average values  
the ``zero momentum Green's functions at finite volume''. 
This set of values can be thought of as 
an element of an infinite vector space. Near the fixed point, we can use the 
eigenvectors of the linearized transformation as a basis. 
As far as we are close to the fixed point, the 
average values of the powers of the {\it rescaled} total field stay 
approximately unchanged after one transformation. 
However at each iteration, the components in the eigendirections are 
multiplied by the corresponding eigenvalue. In the following, we 
discuss the case where there is only one relevant direction,
in other words, only one eigenvalue larger than 1.
We call this eigenvalue $\lambda$. 

After repeating the renormalization group transformation $n$ times,
we have replaced $b^{Dn}$ sites by one site
and associated a block variable with it. If at this point we
neglect the interactions among the blocks of size 
$b^{Dn}$ and larger, we can calculate the finite 
volume volume Green's functions. For the sake of definiteness, we 
only discuss the case of the two point function. In the statistical
mechanics language, the zero-momentum two point function is called the
magnetic susceptibility. We define the finite volume susceptibility 
$\chi_n$ as the average value of the square of the sum of all the 
(unrescaled) fields
divided by the number of sites $b^{Dn}$.

From the above discussion, we estimate 
\begin{equation}
\chi_n\simeq b^{n(2-\eta)}(K_1+K_2\lambda^n(\beta_c-\beta))
\label{eq:chi}
\end{equation}
The power of $b$ comes from rescaling back two powers of the fields to 
their original values, together with the division by the number of sites.
The constant $K_1$ is the constant value of the average of the square of 
the {\it rescaled} sum of the fields
at the fixed point. The constant $K_2$ depends on the way the critical 
hypersurface is approached when $\beta$ is varied close to $\beta_c$.

We want to emphasize that 
Eq. (\ref{eq:chi}) is valid only if the linearization procedure is 
applicable, in other  words if $\lambda^n(\beta_c-\beta)<<1$. 
On the other hand, when $n$ reaches some critical value $n^\star$ such that
\begin{equation}
\lambda^{n^\star}(\beta_c-\beta)\simeq 1\ ,
\label{eq:nstar}
\end{equation}
non-linear effects become important
and the sign of $(\beta_c-\beta)$ becomes important. 
In the following we will consider exclusively the case of the 
symmetric phase which is simpler.
For a discussion in the case of the broken symmetry phase  ($\beta>\beta_c$),
the reader may consult Ref. \cite{hyper}.
If $\beta<\beta_c$, the value of $\chi$ starts stabilizing when $n$ gets 
of the order of $n^\star$. 
As a consequence,
\begin{equation}
\chi_{\infty}\sim b^{n^{\star}(2-\eta)}\sim(\beta_c-\beta)^{-\gamma}\ ,
\end{equation}
with
\begin{equation}
\gamma=(2-\eta){\rm ln}b/{\rm ln}\lambda \ .
\end{equation}
When $n$ gets larger, the
high-temperature fixed point is reached rapidly. This fixed point is completely
attractive. The irrelevant directions are manifested by volume effects 
decreasing like $b^{-n2}$. A model calculation of these 
effects can be found in Ref. \cite{finite}. 

For most models studied in the literature, 
the exact rescaling factor $s=b^{(2+D-\eta)/2}$ 
is not known with
perfect accuracy. For instance in a Monte Carlo calculation for
values of the couplings minimizing the subleading corrections,
the values of $\eta$ obtained (for two-components in 3 dimensions) 
is\cite{eta} 0.0381 with errors of order 2 in the last quoted digit.
Since at the end of the calculation we are 
rescaling back to the original field variables, this uncertainty 
would not affect the physical results provided that
we were able to carry the integrations exactly.
If the integrations are carried numerically, the arithmetic operations 
are performed differently and choosing a sample of values for the 
rescaling factor $s$ close to the approximate values can be used to
obtain a statistical sample of the numerical errors.

We will thus consider values of 
the rescaling factor $s$ of the form $s_{exact}+\delta$, where 
$s_{exact}$ is the value for which there is an exact fixed point. In some 
sense $\delta$ can be seen as the ``seed'' of a random number generator.
We will now perform sample calculations for two models 
and compare our results with 
the results obtained from a set of seeds for a random number generator 
designed to
produce approximate Gaussian distributions.

\section{Model Calculations}
\label{sec:models}

In this section, we describe in detail the three 
procedures used to obtain the data analyzed in the following
sections. We discuss the hierarchical
model\cite{dyson} (subsection A), a simplified version \cite{toy} 
of it (subsection B) and a random number
generator which represents our naive expectations for the two 
other cases (subsection C). 

\subsection{The hierarchical model}
\label{subsec:hmodel}

The choice of the hierarchical model allows easy calculations
with a controllable accuracy. In order to avoid repetitions, we 
refer the reader to Ref. \cite{finite} for a more systematic 
presentation. 
The main interest of this model is that only the local potential
(or equivalently, the part of the measure which factorizes into 
a product of identical local functions which are called 
``the local measure'' later)
is affected by the RG transformation while the ``kinetic'' part is 
left invariant. 

The block-spin 
procedure can then be summarized by a simple integral 
formula (Eq. (2.2) in Ref. \cite{finite}) for the local measure. 
Taking the Fourier transform and rescaling the sum of the fields 
in the block by 
an {\it arbitrary} rescaling factor 
$1/s$ one obtains \cite{finite} the recursion relation
\begin{equation}
R_{n+1}(k) = C_{n+1}
exp(-{1\over2}\beta({c\over4}s^2)^{n+1}{\partial^2\over\partial
  k^2})(R_n({k\over s}))^2 \ .
\label{eq:rec}
\end{equation}
The parameter $c$ is set equal to $2^{1-2/D}$
in order to approximate a $D$-dimensional 
model with nearest neighbor interactions. In the following the value $D=3$
will be used.
 
We fix the normalization constant $C_n$ is such way that
$R_n(0)=1$. $R_n(k)$ has then a direct probabilistic interpretation. 
If we call $M_n$ the total field $\sum\phi_x$ inside blocks
of side $2^n$ and $<...>_n$ the average calculated without taking into
account the interactions among different blocks of this size
find
\begin{equation}
R_n(k) = \sum_{q=0}^{\infty}{(-ik)^{2q}\over 2q!}{<(M_n)^{2q}>_n\over s^{2qn}}
\end{equation}
We see that the Fourier transform of the local measure after $n$
iterations generates the zero-momentum Green's functions calculated
with $2^n$ sites. In particular, we are interested in calculating the 
finite volume susceptibility
\begin{equation}
\chi_n ={<(M_n)^{2}>_n\over 2^{n}}= -2a_{n,1}({s^2\over2})^n \ .
\label{eq:resc}
\end{equation}

As far as we are only interested in the calculation of  $<(M_n)^{2q}>_n$,
the choice of $s$ is a matter of convenience. For the calculations in
the high temperature phase (symmetric phase) not too close to the
critical points, or high temperature expansions the choice $s=\sqrt2$
is natural \cite{high}. On the other hand, the the choice
of rescaling factor $s=2c^{-1/2}$ makes the explicit dependence on $n$ 
in Eq. (\ref{eq:rec}) 
disappear. For any other value of $s$, the map is somehow analogous to 
a differential equation with explicitly time-dependent coefficients.
What we call ``the RG transformation'' in section \ref{sec:strat},
is Eq. (\ref{eq:rec}) with $s=2c^{-1/2}$. 
It corresponds to the values $b=2^{1/D}$, $\eta=0$ and $s=b^{(2+D)/2}$
of the parameters defined 
in section \ref{sec:strat}. For this value of $s$, Eq. (\ref{eq:rec}) has 
non-trivial fixed point\cite{wittwer} which seems to be unique\cite{gam3}.

We have calculated the susceptibility for the Ising measure ($R_0(k)=cos(k)$)
with $D=3$ (i.e. $c=2^{1/3}$) and 
$\beta=\beta_c-10^{-8}$. The calculations have been performed using double 
precision Fortran. Using the approximate error given in Eq. (\ref{eq:reler}), 
we see that 8 out of the 16 significant digits of $\chi$ should 
be correct for $\beta=\beta_c-10^{-8}$. Since we are in the high-temperature 
phase, $\chi_n$ stops growing after approximately $n^{\star}\simeq 52$ 
iterations (see section \ref{sec:strat} and Ref. \cite{gam3} for details).
The 16 digits  of $\chi_n$ are completely stabilized after $n=140$ iterations.
We call this stable value $\chi(s)$ where $s$ refers to the value of $s$ in 
Eq. (\ref{eq:rec}) used for the numerical calculation.
The calculations have performed using dimensional approximations of degree
$l_{max}$:
\begin{equation}
R_n(k) = 1 + a_{n,1}k^2 + a_{n,2}k^4 + ... + a_{n,l_{max}}k^{2l_{max}}\ ,
\end{equation}
for which the recursion
formula for the $a_{n,m}$ reads :
\begin{equation}
a_{n+1,m} = {
{\sum_{l=m}^{l_{max}}(\sum_{p+q=l}a_{n,p}a_{n,q}){[(2l)!/(l-m)!(2m)!]}({c/4})^l[-(1/2)\beta]^{l-m}}\over{\sum_{l=0}^{l_{max}}(\sum_{p+q=l}a_{n,p}a_{n,q}){[(2l)!/l!]}{(c/4)^l}[-(1/2)\beta]^l}}
\end{equation}
We found that the 16 digits of $\chi_n$ were completely stabilized 
for $l_{max}=50$.

We have used the set of values $s=2/\sqrt{c}\pm m\times 10^{-8}$
with $m=1,\dots 10^5$. 
We recorded the difference of $\chi(s)$ defined above 
with respect to the accurate value
$\chi=5.2316268857268\times 10^{10}$. This value was obtained by performing
the calculation using higher precision arithmetic (namely 30 digits). 
We have checked that this accurate result is 
insensitive to changes in $s$. 
The results are discussed in the next sections.

\subsection{A simplified model}
\label{subsec:tmodel}

As noticed in Ref. \cite{finite},
for $\beta<\beta_c$ and $n>>n^\star$ 
Eq. (\ref{eq:rec}) imples the approximate behavior
\begin{equation}
\chi_{n+1}\simeq \chi_n +(\beta/ 4) ({c/2})^{n+1}\chi_n^2\ .
\end{equation}
This map has been studied \cite{toy} on its own in the rescaled form
\begin{equation}
h_{n+1}=(c/2)h_n+(1-c/2)h_n^2 \ .
\label{eq:hmap}
\end{equation}
This map can be seen as a drastically simplified version of the 
hierarchical model. One can check that if $0\leq h_0 \leq 1$, 
lim$_{n->\infty}(2/c)^n h_n$ is finite. This limit plays the role of
the susceptibility in the following and can be calculated accurately
by combining dual expansions \cite{toy}.

Eq. (\ref{eq:hmap}) has an unstable fixed point at $h=1$ with 
eigenvalue $\lambda=2-c/2$.
We have required $\lambda=1.427$, approximately as for the 
hierarchical model with $D=3$, 
in order to keep the value of $n^\star$ the 
same. Consequently, the value of $c$ used in Eq. (\ref{eq:hmap})
is {\it not} the same as the value of $c$ for the
hierarchical model with $D=3$.

We have introduced the rescaling factor $s$ through the redefinition
\begin{equation}
a_n\equiv(s^2c/4)^{-n}h_n \ ,
\label{eq:aresc}
\end{equation}
in terms of which the map becomes 
\begin{equation}
a_{n+1}=(2/s^2)a_n+(1-c/2)(s^2c/4)^{n-1}a_n^2
\label{eq:amap}
\end{equation}
After calculating, $a_n$ one can always return to $h_n$ using 
Eq. (\ref{eq:aresc}). If we had the chance to use exact arithmetic
the expression would be independent of $s$.

We have performed calculations for  $h_0=a_0=1-10^{-8}$.
We found that 150 iterations were sufficient to stabilize 
lim$_{n->\infty}(2/c)^n h_n$. We have calculated  this value for
$s=2/\sqrt{c}\pm m\times 10^{-8}$ with $m=1,\dots 10^5$, as in the 
previous case. We have then subtracted the more accurate 
value
\begin{equation}
{\rm lim}_{n->\infty}(2/c)^n h_n=3.842965603774557\times 10^{12} \ .
\end{equation}
obtained and checked exactly with the same procedure 
as in the previous subsection.

\subsection{A model with gaussian distribution}
\label{subsec:rng}

In order to provide a comparison of the errors distributions of the 
two previous models, we have also generated $10^5$ numbers using a method
designed to give a gaussian distribution. 
Since in the two
previous cases we have approximately $n^\star=52$ random processes
before the value recorded starts stabilizing, we have added 52 
random numbers. These random numbers have been produced by
using repeated multiplication by a
large number ($7^5$) followed by a reduction modulo 1 (in other words, we drop 
the integer part). 
This procedure is inspired by results reviewed in Ref. \cite{lewis}.
Iterating this procedure, we generate a sequence of random 
numbers which we expect to be uniformly distributed
between 0 and 1. 
In order to get numbers distributed between -1 and 1, we multiply each number
by 2 and subtract 1.
Finally, in order to get numbers
approximately of the same order as the numbers of the other two sets, 
we have multiplied the final sum of 52 random numbers 
by 1000. The final results depends only 
on the initial number provided (the ``seed''). We have repeated this 
calculation for $10^5$ values of the seed between 0 and 1.

The above discussion can be summarized as follows.
We iterate 52 times the map
\begin{equation}
\alpha_{n+1}=7^5\alpha_n \ {\rm Modulo}\  1 ,
\label{eq:alpha}
\end{equation}
and then calculate
\begin{equation}
X_j=1000\times \sum_{n=1}^{52}(2\alpha_{n}-1)\ .
\label{eq:aver}
\end{equation}
The sub-index $j$ corresponds to different values of the seed 
$\alpha_0$. The data analyzed in the next sections corresponds the 
choice
\begin{equation}
\alpha_0(j)=j/100005.23 \ ,
\end{equation}
with 
$j=1,\dots 10^5$. The choice of the denominator is motivated by the
fact that we want a spacing between successive initial values 
slightly smaller than $10^{-5}$. The decimal points (.23) have been
adjusted empirically in order to get a good uniformity 
of in subsamples (this 
question is discussed in section \ref{subsec:uni}).

The map of Eq. (\ref{eq:alpha}) is designed to  provide a sample from a 
variable $\alpha$ which we expect to be 
uniformly distributed between 0 and 1.
If this is the case, we should have
$\langle (2\alpha-1)\rangle=0$ and Var$(2\alpha-1)=1/3$.
We use the common notation
Var$(A)=\langle(A-\langle A\rangle)^2\rangle$. 
If we now define, 
$X=1000\times \sum_{n=1}^{52}(2\alpha_{n}-1)$, 
we expect
\begin{eqnarray}
\langle X \rangle&=&0\ ; \\ \nonumber
{\rm Var}(X)&=&(1000)^2\times (52/3)\simeq (4163)^2 \ .
\label{eq:unipred}
\end{eqnarray}
These predictions will be tested in section \ref{sec:moma}.

\section{Error distributions} 
\label{sec:distr}

In order to get a first idea about what can be done to characterize the 
distributions of values obtained by the procedures described in the 
previous section, we first 
consider the hierarchical model and 
display a subsample of 1000 data points 
for $s$ below $s_{exact}=2/\sqrt{c}$ and 1000 data points 
for $s$ above this value. 
The selection of the subsample was done  by taking one out of every 100
values out of the original sample. 
More precisely, the integer $m$ used in the parametrization of $s$ 
given in subsection \ref{subsec:hmodel},
takes only values which are multiples of 100. 
The distribution of values is shown in Fig.
\ref{fig:deltachi}. 

One immediately realizes that the distribution is not symmetric
about $s=2/\sqrt{c}$. If $s>2/\sqrt{c}$, 
the values of $\chi$ are more spread than if $s<2/\sqrt{c}$.
An histogram can be obtained by dividing the data points 
into ``horizontal bins'' of equal vertical height and counting the number 
of data points in each bin. This procedure has been followed for 50,000
data points with $s>2/\sqrt{c}$ and 50,000
data points with $s<2/\sqrt{c}$. Each of the two sets is obtained by taking
every other point out of the data 
for the hierarchical model described in the previous section.
With this procedure the total number of data points is still $10^5$
and the statistical properties can be easily compared with the 
other samples of $10^5$ data points. 
The histogram is displayed in Fig. \ref{fig:habbar}.

The solid line is obtained by plugging the estimated 
mean and standard deviation (discussed in the next section) in 
a gaussian probability distribution. In the rest of this section, 
the terminology ``Gaussian fit'' refers to this procedure.
One sees that there are large deviations
from the Gaussian fit. Given the size of the sample it is very unlikely 
that the deviations can be interpreted as statistical fluctuations. 

For comparison, we give in Fig. \ref{fig:lewisbar} 
an histogram of the distribution of $10^5$
data points obtained from the random number generator discussed 
in section \ref{subsec:rng}.
The features of the distribution can be seen better by plotting the 
logarithm of the number of data  points in each bin. 
In this semi-log plot, the fit corresponding to a Gaussian
distribution is simply a parabola. 
Also, we will study separately, the data points with 
$s<2/\sqrt{c}$ and $s>2/\sqrt{c}$ since from Fig. \ref{fig:deltachi}
they have manifestly different standard deviations.
The results for the two sets of $10^5$ data points described in 
subsection \ref{subsec:hmodel} for 
the hierarchical model are shown in Fig. \ref{fig:hbar}.

These distributions are roughly of rectangular shape with slow
modulations on the ``flat'' part. They can be compared with a 
similar graph for the random number generator (Fig. \ref{fig:lewbar}).
On this graph, the departure from the parabolic behavior is 
barely visible except in the tails. Obviously, bins with zero data
points cannot be shown in such a graph.

Finally, we have displayed the results 
for the two sets of $10^5$ data points described in 
subsection \ref{subsec:tmodel} for 
the toy model (Fig. \ref{fig:tbar}).
One sees that significant deviations 
from the Gaussian fit appear in the tails. The fact that the
probability distribution falls more rapidly in this region is indicative
of a probability distribution of the form exp$(-ax^2-bx^4)$ with $a$ and $b$
positive. This relative simplicity suggests that we have a better chance
to fully understand this simplified example.

\section{Moment Analysis}
\label{sec:moma}

\subsection{Moment estimators}

A probability distribution is characterized by its moments.
We would like to estimate the first moments of the distributions described
in the previous sections {\it assuming} that they are sample of a unique 
probability distribution.
Let $X_i$ with $i,\dots N$ be a set of independent and identically distributed
(i.i.d.) random variables with moments
\begin{eqnarray}
\langle X_i \rangle &=& \mu\ ; \\ \nonumber
\langle (X_i -\mu)^r \rangle&=& \mu_r\ , r\geq 2 \ ,
\label{eq:mom}
\end{eqnarray}
identical for any $i$. In particular, $\mu_2=\sigma^2$ is the 
variance or the square of the standard deviation $\sigma$.
We define the estimators
\begin{eqnarray}
\nonumber
\widehat{\mu}&=&(1/N)\sum_1^N X_i\ ;  \\ 
\label{eq:est}
\widehat{\mu_r}&=&(1/N)\sum_1^N (X_i -\widehat{\mu})^r\ , r\geq 2\ .
\end{eqnarray}
Using the hypothesis that the random variables are independent, we can 
factor the powers of a given $X_i$ and calculate its average independently.
For instance, if $i\neq j$, $\langle X_iX_j\rangle=
\langle X_i\rangle \langle X_j \rangle$. Using the hypothesis that the
random variables are identically distributed, we can use Eq. (\ref{eq:mom})
to express these expectations in terms of the common values of 
their moments. Using these rules, one obtains
\begin{eqnarray}
\langle \widehat{\mu}\rangle&=&\mu\ ; \\
\langle\widehat{\mu_r}\rangle&=&\mu_r+O(1/N)\ , r\geq 2 \ .
\end{eqnarray}
The biases of the $\mu_r$ are of order $1/N$.
It is not difficult to remove the bias, however since we will 
work with large samples, the corrections are very small.

In Table I, we give the estimated values of $\mu$ and $\sigma$
for the distributions discussed before.
The abbreviation a. is short for ``above'' which means the $10^5$ points
with $s>2/\sqrt{c}$ for the hierarchical model (H.M.) ot the toy 
model (T.M.). Similarly b. is short for ``below'' ($s<2/\sqrt{c}$), 
while a.+b. means ``above and below'' which is short for $10^5$ points
obtained by taking every other value in the ``above'' and ``below'' set
as explained at the beginning of section \ref{sec:distr}.
Finally, ``R.N.G.'' is short for the random number generator described in
subsection \ref{subsec:rng}. These notations will be used again 
in the following.
We emphasize that 
a.+b. is {\it not} the union of a. and b. (so $\mu_{a.+b.}\neq 
(1/2)(\mu_{a.}+\mu_{b.}$))
and that all the samples have $10^5$ data points.

Initially we expected that $\mu$ would be of order $\sigma/\sqrt{N}$. However,
this is only the case for the random number generator.
In all the other cases, $\mu$, which we remind is the average difference 
with respect to the accurate value, is of the same order of magnitude as 
 $\sigma$. In other words, it seems impossible to get rid of the 
errors by using large statistics! 

In order to check the compiler-dependence of our results, we have repeated 
the calculations of the toy model in C, Mathematica and a different 
version of Fortran. We found that $\mu$ was only affected by less than 
10 percent. On the  other hand, $\sigma$ varied much more 
significantly.

Note also that in the case of the R.N.G., the estimated value of 
$\sigma$ is close to the result (4163) obtained in Eq. (\ref{eq:unipred})
by assuming 
that the $\alpha_l$ are uniform over the interval
$[0,1]$.

As expected, the $\sigma$ ``above'' and ``below'' are significantly different.
It is very unlikely that the data 
``above and below'' is a sample from a unique distribution. 
More generally, one could study subsamples and check that 
the fluctuations are compatible with their size.
This is the topic of the next subsection.

\subsection{Fluctuations and uniformity}
\label{subsec:uni}

In the previous subsection, we have used moments estimators which
are in good approximations unbiased. However, it is not clear that 
the sample comes from a unique distribution.
One way to tackle this question is to consider the estimators in
many subsamples and decide if the the fluctuations 
of the estimated values of the moments in the subsamples 
are compatible
with the variance of the estimator which we now proceed to estimate.

Using again the hypothesis of independent and identical distributions, we 
find in leading order in $1/N$ (in other words, 
up to $O(1/N^2)$ corrections) that:
\begin{eqnarray}
Var( \widehat{\mu})&=&(1/N)(\mu_2)\ ; \\ 
Var( \widehat{\mu_2})&=&(1/N)(\mu_4-\mu_2^2)\ ; \\ 
Var( \widehat{\mu_3})&=&(1/N)(\mu_6-\mu_3^2+9\mu_2^3-6\mu_2\mu_4)\ ; \\ 
Var( \widehat{\mu_4})&=&(1/N)(\mu_8-\mu_4^2+16\mu_2\mu_3^2-
8\mu_3\mu_5)\ . 
\end{eqnarray}
The proof of these results can be found for instance in Ref. \cite{kendall}.

We considered the seven sets of $10^5$ data points discussed in 
the previous 
subsection. 
Each set has been divided into 
$m$ subsets with $N_B$ data points. Obviously,
$N=m\times N_B$. The partition has been done 
by putting together successive values of $s$.
If one refers to Fig. \ref{fig:deltachi}, the subsets are vertical partitions.
We call $\widehat{\mu_{r,S}}$ (or $\widehat{\mu_S}$) 
the estimator of $\mu_r$ (or $\mu$) in the 
$S$-th subsample. These are defined as in Eq. (\ref{eq:est}), except for the 
fact that $N$ needs to be replaced by $N_B$. In order to avoid 
confusion, we have used a subscript $T$ (short for Total) to designate 
the estimators in the whole sample. We have studied the differences 
between the subsample estimates and the whole sample estimates. 
In order to get comparable answers, we have expressed these differences in
``natural units''. These units should be such that if we had
a sample from a unique distribution, all the fluctuations would be of order 1.
From the expressions of the variances, ones sees that
$\sigma^r/\sqrt{N_B}$ are natural units 
for the fluctuations of
$\widehat{\mu_{r,S}}$. In practice, we have to replace $\sigma$ by its 
estimated value.
As an illustration, we have taken three sets previously abbreviated 
as H.M.a.+b., T.M.a.+b. and R.N.G. and divided them into $m=400$ sub-bins.
These sub-bins are ordered according to increasing values of $s$. 
For illustration,
we have displayed the fluctuations of the second moment in natural units.
The results are shown in Fig. \ref{fig:fluct}.

One sees that for the first two sets, there is a
 ``jump'' at $s=2/\sqrt{c}$ that is 
significantly larger than the other fluctuations. 
Note that we have checked  that both a.+b. sets 
had their sub-bins ordered in the same way.
Besides these jumps, the fluctuations appear to be of much more ``normal''
size. In order to make this visual impression quantitative, we have 
defined the average of the square of the fluctuations in the subsamples:
\begin{eqnarray}
\nonumber
U_1&=&(1/m)\sum_{S=1}^m(\widehat{\mu}_S-\widehat{\mu_T})^2\ ; \\ 
U_r&=&(1/m)\sum_{S=1}^m(\widehat{\mu_{r,S}}-\widehat{\mu_{r,T}})^2\ .
\end{eqnarray}
If the sample comes from a single distribution, 
we expect these quantities to be equal to the variance 
of the corresponding estimator with
fluctuations of order $1/\sqrt{m}$. 
We have thus defined some ``indicators of uniformity'' as
\begin{eqnarray}
\nonumber
p_1=U_1/{\rm Var}(\widehat{\mu}_S)\ ; \\
p_r=U_r/{\rm Var}(\widehat{\mu_{r,S}})\ .
\end{eqnarray}
If the sample comes from a single distributions, the $p_i$ should 
be close to 1, while a large value indicates the 
presence of several distributions. The values of these indicators 
are given for the seven sets of data points previously described in Table II.

First, one notices that the R.N.G. data has all the $p_i$ within less than 
ten percent of 1. On the other hand, both a.+b. sets have large
values. For the toy model, the separation into above and below
is sufficient to get rid of these large values.
The values for T.M.a. and T.M.b 
are within 20 percent of 1. 
This seems in good approximation compatible with 
a single distribution. This quantitative analysis confirms the visual 
impression that one gets from Figs. \ref{fig:hbar} and \ref{fig:tbar}.
On the other hand, 
it is possible to 
further ``resolve'' the a. and
b. data for the hierarchical model. 

In the case of the hierarchical model, the local analysis of the sub-samples
shows that the rapid variations of the mean has ``large-scale'' tendencies.
Namely we observed 
regions of $s$ where in average, the mean grows linearly when $s$
increases followed by regions where it decreases. 
This behavior is presently under study.

\subsection{Departure from Gaussian behavior}

In section \ref{sec:distr}, we noticed that the distributions for the toy
model a. and b. were more sharply peaked than a Gaussian distribution. 
This feature can be analyzed more quantitatively by estimating the 
skewness coefficient $\mu_3/\mu_{2}^{3/2}$ and the coefficient of
kurtosis ${\mu_4/\mu_2^2}-3$. 
For a Gaussian distribution, both coefficients are zero. 
These estimations of these coefficients for the data sets previously
considered are given in Table III.

For the T.M. a. and b., both coefficients are 
an order of magnitude larger than the corresponding
coefficients for the R.N.G. . Larger deviations are observed for
the other sets reflecting the lack of uniformity of these sets
of data.

In summary, the above table supports the general statement that
the numerical errors appearing in RG calculations are non-Gaussian.
We will now attempt to explain this general feature with a simple model.

\section{A model for non-Gaussian errors}
\label{sec:moder}

One reason to be inclined to believe {\it a-priori} that errors distributions
should be Gaussian is the famous central-limit theorem which 
asserts that the average 
of $N$ i.i.d. random variables approaches a Gaussian distribution when 
$N$ becomes infinite. 
In this section,
we give an heuristic derivation of this theorem and then present 
a mechanism by which we can avoid ending up with Gaussian distributions.

\subsection{The central limit theorem}
\label{subsec:central}

We first review the central-limit theorem in terms of a moment
analysis. Let 
$\alpha_j$ $j=1,\dots n$ be a set of i.i.d. random variables 
with arbitrary moments ($\mu, \mu_2=\sigma^2,\mu_3 \dots$).
We define 
\begin{equation}
Z=\lbrack(1/n)\sum_{j=1}^n(\alpha_i-\mu)\rbrack /(\sigma/\sqrt{n})\ .
\end{equation}
It is clear that $Z$ has $\mu=0$ and $\sigma=1$.
The central-limit asserts that
\begin{equation}
{\rm Lim}_{n->\infty}P(Z)=(1/\sqrt{2\pi})e^{-Z^2/2}\ .
\end{equation}
This theorem can be proven by showing that the 
moments of $Z$ coincide with the moments of a normal Gaussian : 
\begin{eqnarray}
\label{eq:gm}
\mu_{2r}&=&(2l-1)!!\ ; \\
\mu_{2r+1}&=&0\ .
\end{eqnarray}

The calculation of the expected values of $Z^q$ goes 
as follows. Each term of $Z^q$ is a product of $q$ 
terms of the form $(\alpha_i-\mu)$.
The expected value can be factored into products involving the 
same $i$ and then expressed in terms of the moments. Since
$\langle \alpha_i-\mu\rangle=0$, each $i$ should be repeated at least twice.
For $Z^3$, the three $i$ should be the same or one will be ``left alone'' 
(which would imply a vanishing average). Since the three indices have to be 
the same and since there are $n$ indices
we get a combinatoric factor $n$ which is insufficient to overcome the 
$n^{-3/2}$ appearing in $Z^3$. A similar reasoning 
in the case of $Z^4$, shows there are three ways to arrange four indices into 
two pairs having the same index (assuming that the index of the two 
pairs are distinct). If we require that the four indexes are the same, it costs
an additional $1/n$ suppression. In summary, when $n$ becomes large
\begin{eqnarray}
<Z^3>=n\mu_3/(\sigma^3 n^{3/2})->0\ ; \\
<Z^4>=(n\mu_4+3n(n-1)\sigma^4)/(\sigma^4 n^{2})->3 \ .
\end{eqnarray}
The argument generalizes easily, and one realizes that the odd moments
vanish and the even moment reduce to the number of unordered pairs,
in agreement with Eq. (\ref{eq:gm}).

\subsection{Evading the central limit}

In order to see how the central-limit can be evaded, we have constructed
a simplified model for the numerical errors. 
In the rest of this section, when we say ``the quantity we are calculating 
numerically'' this means the $R_n(k)$ function 
defined in Eq. (\ref{eq:rec}) 
for the hierarchical model or the quantity $a_n$ of 
Eq. (\ref{eq:amap}) for the toy model. 
While we are near the fixed point, the numerical values appearing in  
these quantities 
which are relevant for our discussion are roughly of order 1.

We first assume that the initial
data ($R_0$ or $a_0$) has an error ${\cal E}_0$ in the unstable direction.
Using the linearized model of section \ref{sec:strat}, we 
see from Eq. (\ref{eq:chi}) 
that as far as we are near the fixed point, 
the quantity we are calculating numerically is of order $K_1$.
After one iteration, ${\cal E}_0$ is amplified by a factor $\lambda$
In  addition, an  error of order $K_1 10^{-16}$ is  made. 
We express our ignorance about the details of the round-off 
errors in terms of a random
variable $\alpha_1$
which takes positive or negative values of order 1 and write 
the new error as $K_1\alpha_1 10^{-16}$.
Putting the two terms
we get
\begin{equation}
{\cal E}_1=\lambda{\cal E}_0+K_1\alpha_1 10^{-16} \ .
\end{equation}
Iterating $n$ times, we get
\begin{equation}
{\cal E}_n=K_1(\sum_{j=0}^n\lambda^{n-j}\alpha_j)10^{-16} \ .
\label{eq:error}
\end{equation}
In order to have compact notations, 
we have introduced a random variable $\alpha_0$ such that
${\cal E}_0=K_1\alpha_0 10^{-16}$. This term is not very important 
for the rest of the discussion, we might have set ${\cal E}_0=0$
and considered errors for a given initial data expressed with 
a finite precision. In that case, the sum in Eq. (\ref{eq:error})
would run from 1 to $n$.

The main difference between the expression of the error of  
Eq. (\ref{eq:error}) and the variable $Z$ introduced in the discussion
of the central-limit theorem, is that the $\alpha_i$ appear with different 
weights. One can then escape the ``pair dominance'' mechanism which 
puts all the individual errors on the same footing.
Instead, the short-distance errors (small $n$)
are more amplified than the 
large-distance errors (large $n$).
Consequently, the total error is likely to inherit some of the features of 
the individual errors. This intuitive picture is confirmed by an 
explicit calculation that we now proceed to explain.

We will calculate the skewness and kurtosis 
associated with
\begin{equation}
Y=\sum_{j=0}^n\lambda^{n-j}\alpha_j \ ,
\end{equation}
assuming that all the $\alpha_i$ are independent and identically distributed.
A detailed study of the errors associated with one iteration of the RG 
transformation for the toy model 
shows that this assumption is not totally correct but that 
it is a reasonable first-order approximation.
A simple calculation shows that after neglecting terms of order 1
compared to $\lambda^n$, we obtain
\begin{eqnarray}
<(Y-<Y>)^3>/(<(Y-<Y>)^2>)^{3/2}& \simeq &
\lbrack(\lambda^2-1)^{3/2}/(\lambda^3-1)\rbrack
(\mu^3/\sigma^{3/2}) \ ; \\
\lbrack<(Y-<Y>)^4>/(<(Y-<Y>)^2>)^2\rbrack-3&\simeq&
\lbrack(\lambda^2-1)^2/
(\lambda^4-1)\rbrack \lbrack(\mu^4/\sigma^4)-3\rbrack \ .
\end{eqnarray}
One sees that $Y$ inherits the non-Gaussian behavior of the $\alpha$'s, 
in the sense that the skewness and kurtosis coefficients of $Y$ are 
proportional to those of $\alpha$.

For $\lambda=1.427$ and 
$\alpha$ uniformly distributed between -1 and 1 (corresponding
to an error of $\pm 1$ in the last significant digit) and which
is roughly what we observed for the toy model, the skewness 
and kurtosis coefficients of $Y$ are 0 and -0.41 respectively.
These numbers are consistent with the 
values obtained for the toy model. A more detailed study shows that if we 
replace the average in Eq. (\ref{eq:aver}) for the random number
generator by a weighted average as in Eq. (\ref{eq:error}), one obtains 
histograms very similar to those of T.M. a. or b. (see Fig. \ref{fig:tbar}).

\section{Conclusions}

We have shown that in RG calculations, perturbations occurring at 
different scales fail to average each other out and that the final result
is most affected by the peculiarities of the short-distance perturbations.
In our study the perturbations are ``anthropomorphic'' in the sense that
they are due to our calculation procedure rather then to natural
phenomena. 
We have also shown that it is not possible to get rid of the 
effects by averaging over calculations at slightly different 
values of the rescaling parameter $s$ (which is not known exactly in most 
realistic
situations).
The only way the errors 
can be reduced to an acceptable level is by using
enough significant digits in
the arithmetic used in the calculations.

The situations is somehow similar to what occurs in chaotic dynamical 
systems. There, the sensitive dependence on the initial conditions,
implies that for time scales 
large compared to the inverse Lyapounov's exponents,
one needs more significant digits than possibly achievable in order to 
keep track of a particular orbit. In the calculations performed here, the
product of the inverse  
Lyapounov exponent (log($\lambda$)) and the time of calculation ($n^{\star}$)
grows only like $(2/\gamma ) {\rm log}(\Lambda/m)$ and one can
obtain reasonably precise calculations of the renormalized mass with
$(\Lambda/m)\sim 10^{17}$ by using only 40 significant digits 
(for $\gamma =1$).
If we are interested in calculating the 
connected higher point functions, more
digits are necessary because some digits are lost in the subtraction 
procedure as explained in Ref. \cite{finite}. In conclusions, we see that
if we are interested in the lowest point connected functions and if we 
have hierarchies of only 17 order of magnitudes, there is no 
serious limitation in our capability to calculate using the RG method.

Finally, it should be noteds that 
non-Gaussian distributions are a common feature in the study 
turbulence \cite{frish}. A simple example is the distribution of 
transverse velocity increments between two points separated by a 
distance $l$ in a turbulent jet, denoted $\delta v_{\perp}(l)$.
Typically, the distribution starts developing ``long tails'' when 
$l$ becomes sufficiently small. In the case studied here, we 
have the opposite effect: our distribution have ``short tails''.
This is due to the fact that the ``microscopic'' 
errors are the individual round-offs which have no tails at all.

\acknowledgments
{This research was supported in part by the Department of Energy
under Contract No. FG02-91ER40664. We thank A. Bhattacharjee and 
J. Scudder for discussions on non-Gaussian distributions.  }

}

\begin{figure}
\vskip50pt
\centerline{\psfig{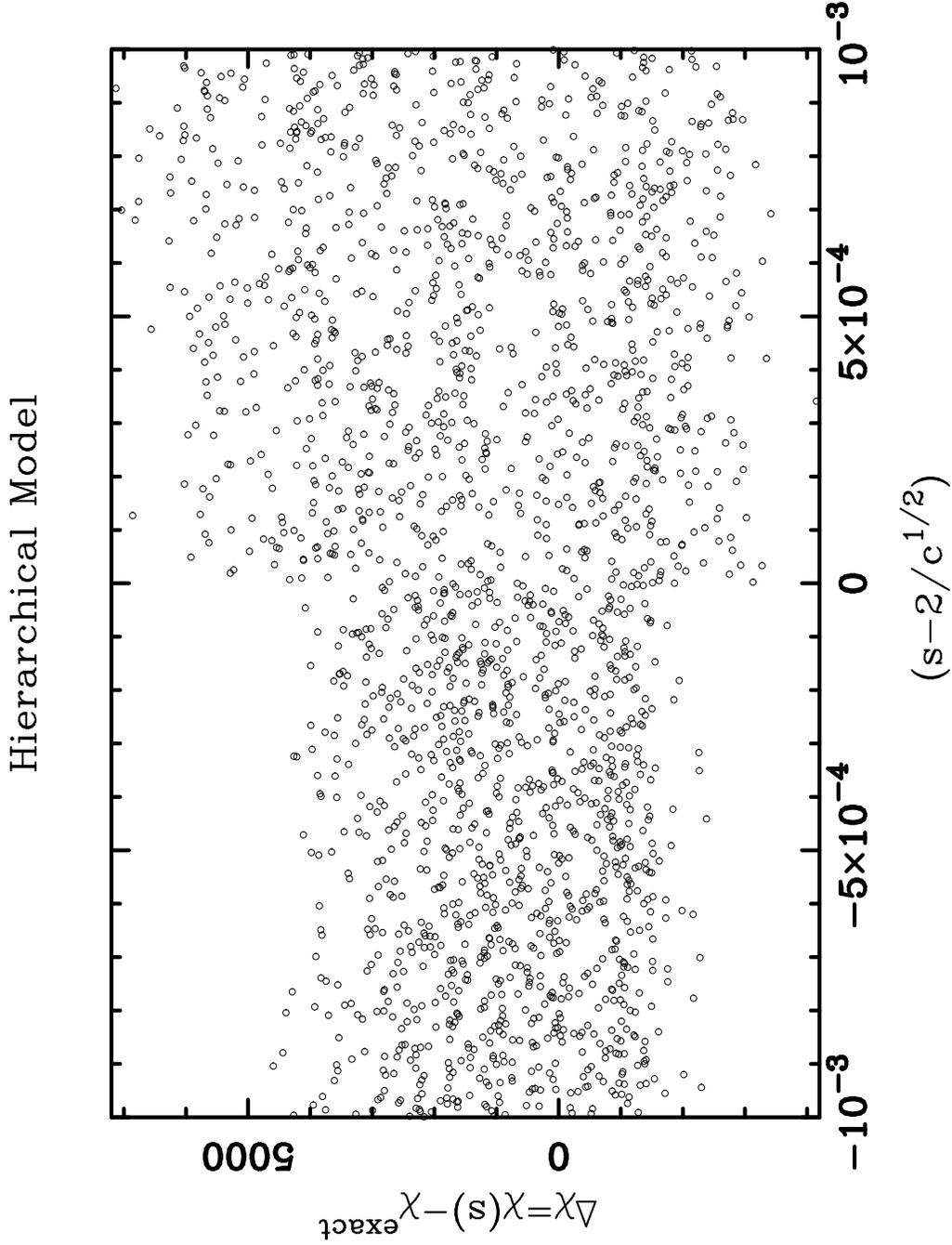}}
\vskip20pt
\caption{Distribution of $\chi(s)-\chi^{exact}$ 
for the hierarchical model
for various values of the rescaling variable $s$.}
\label{fig:deltachi}
\end{figure}

\begin{figure}
\vskip0.4in
\centerline{\psfig{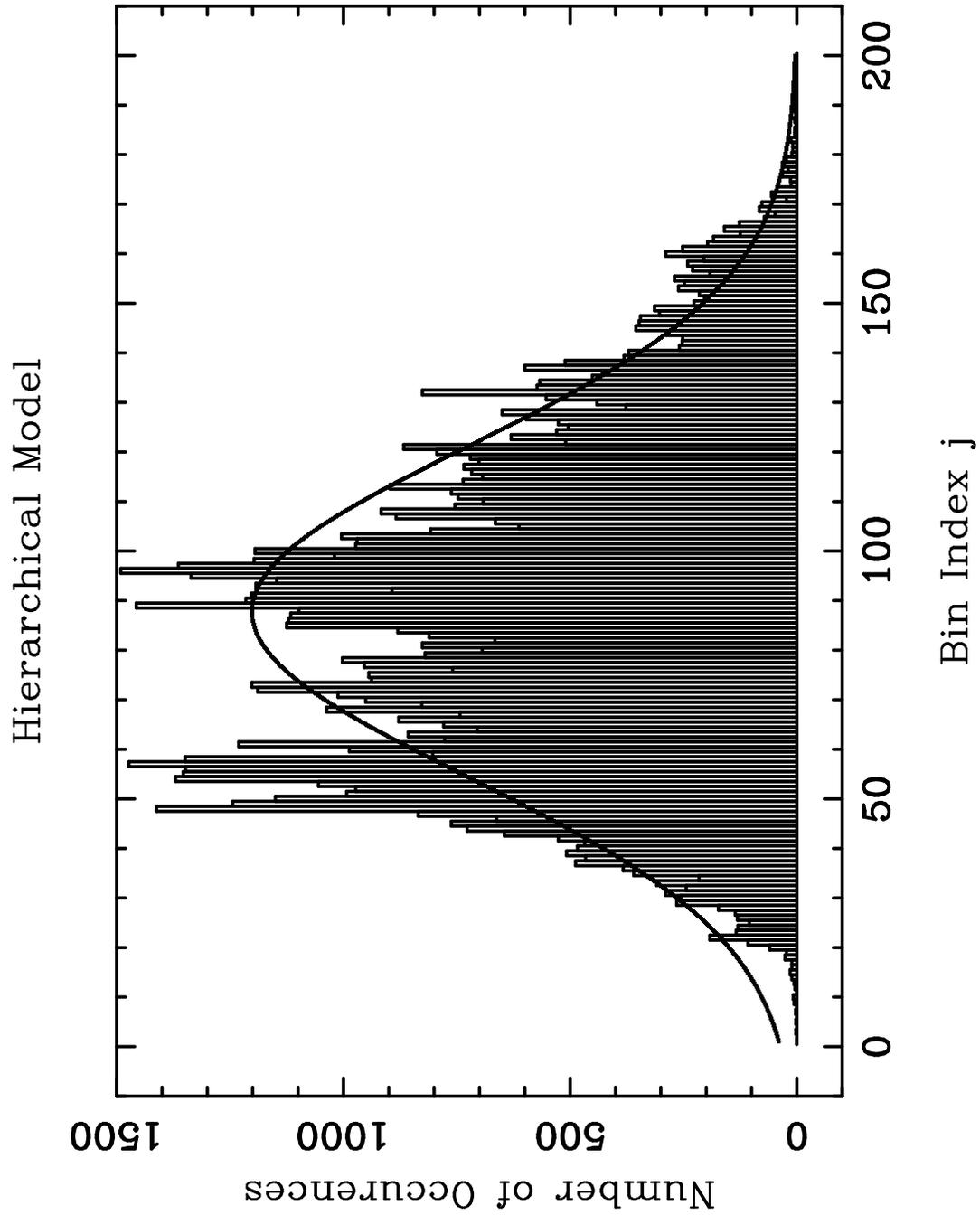}}
\vskip0.4in
\caption{Partition of the $10^5$ values of $\chi(s)$ into 
200 bins as described in the text.
The solid line is a gaussian fit.}
\label{fig:habbar}
\end{figure}

\begin{figure}
\vskip0.5in
\centerline{\psfig{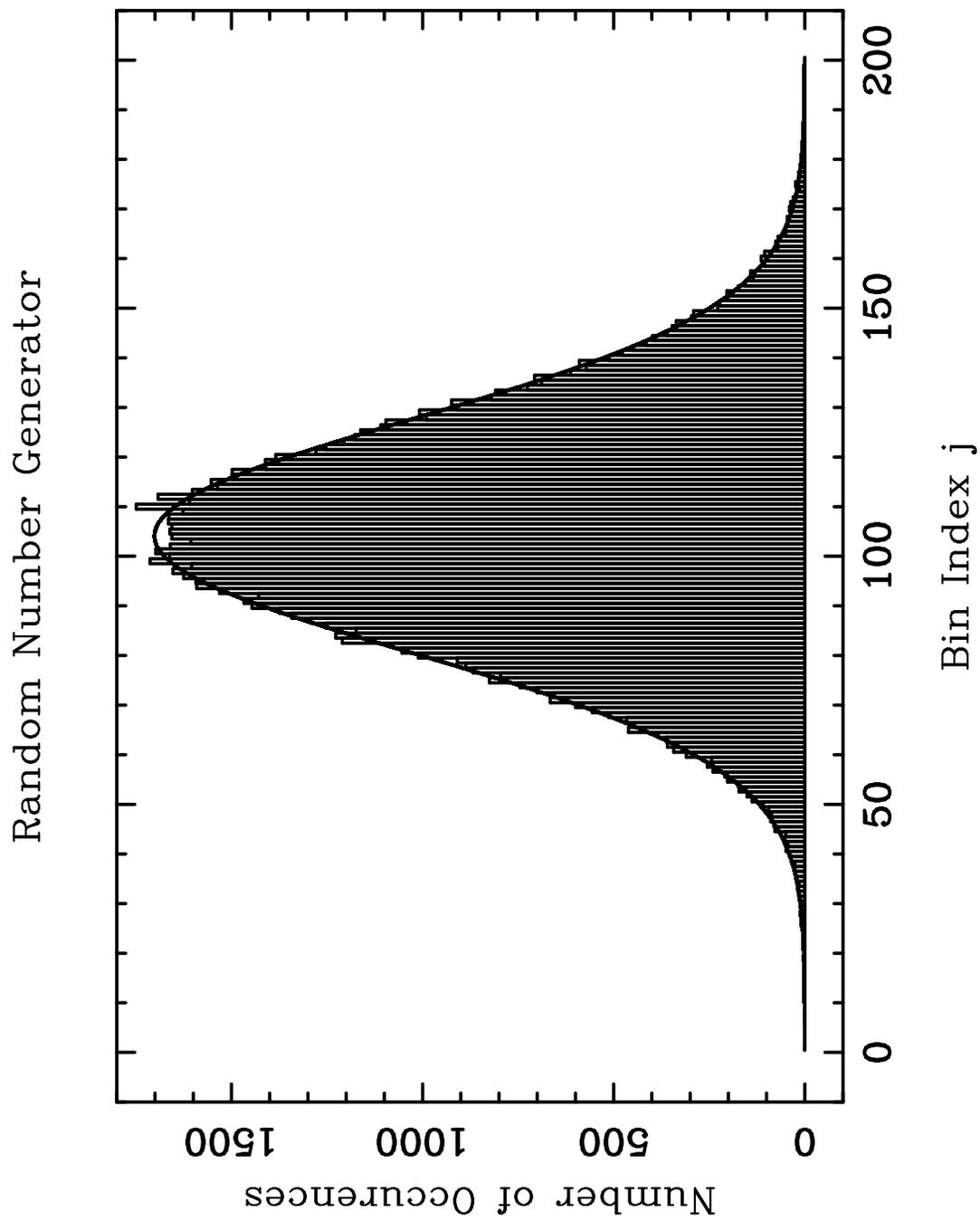}}
\vskip0.5in
\caption{Partition of the $10^5$ values obtained using the 
random number generator into 
200 bins as described in the text.
The solid line is a Gaussian fit.}
\label{fig:lewisbar}
\end{figure}

\begin{figure}
\vskip0.5in
\centerline{\psfig{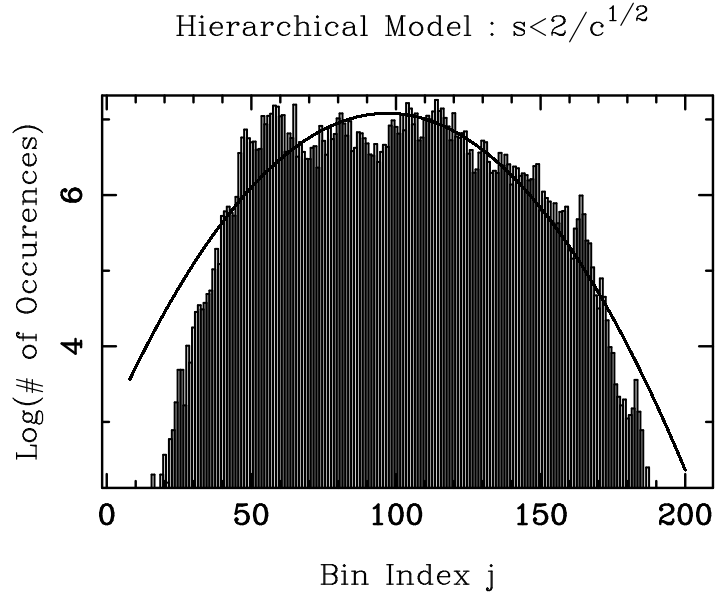}}
\vskip0.5in
\centerline{\psfig{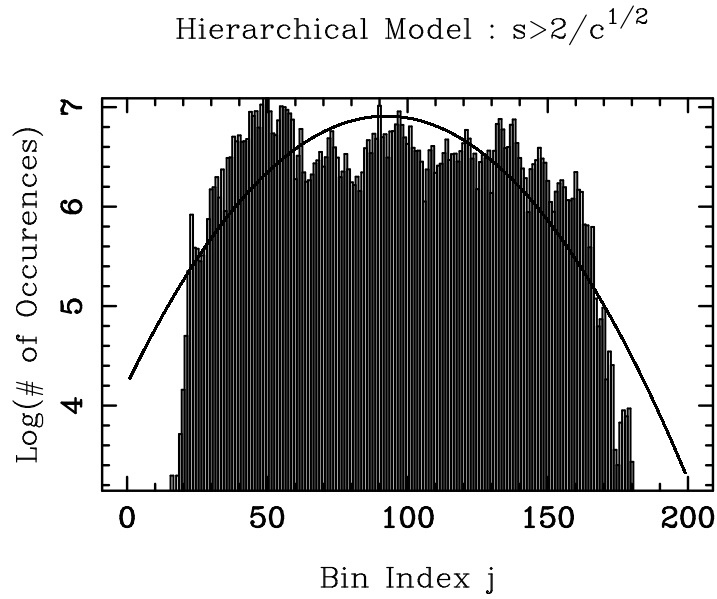}}
\vskip0.5in
\caption{Partition of the $10^5$ values of $\chi(s)$
for $s<2/\sqrt{c}$ and $10^5$
values for $s>2/\sqrt{c}$ into 
200 bins, for the hierarchical model. The logarithm of the number
of data points in each bin is plotted versus the bin number.
The solid parabola is a Gaussian fit.}
\label{fig:hbar}
\end{figure}

\begin{figure}
\vskip0.5in
\centerline{\psfig{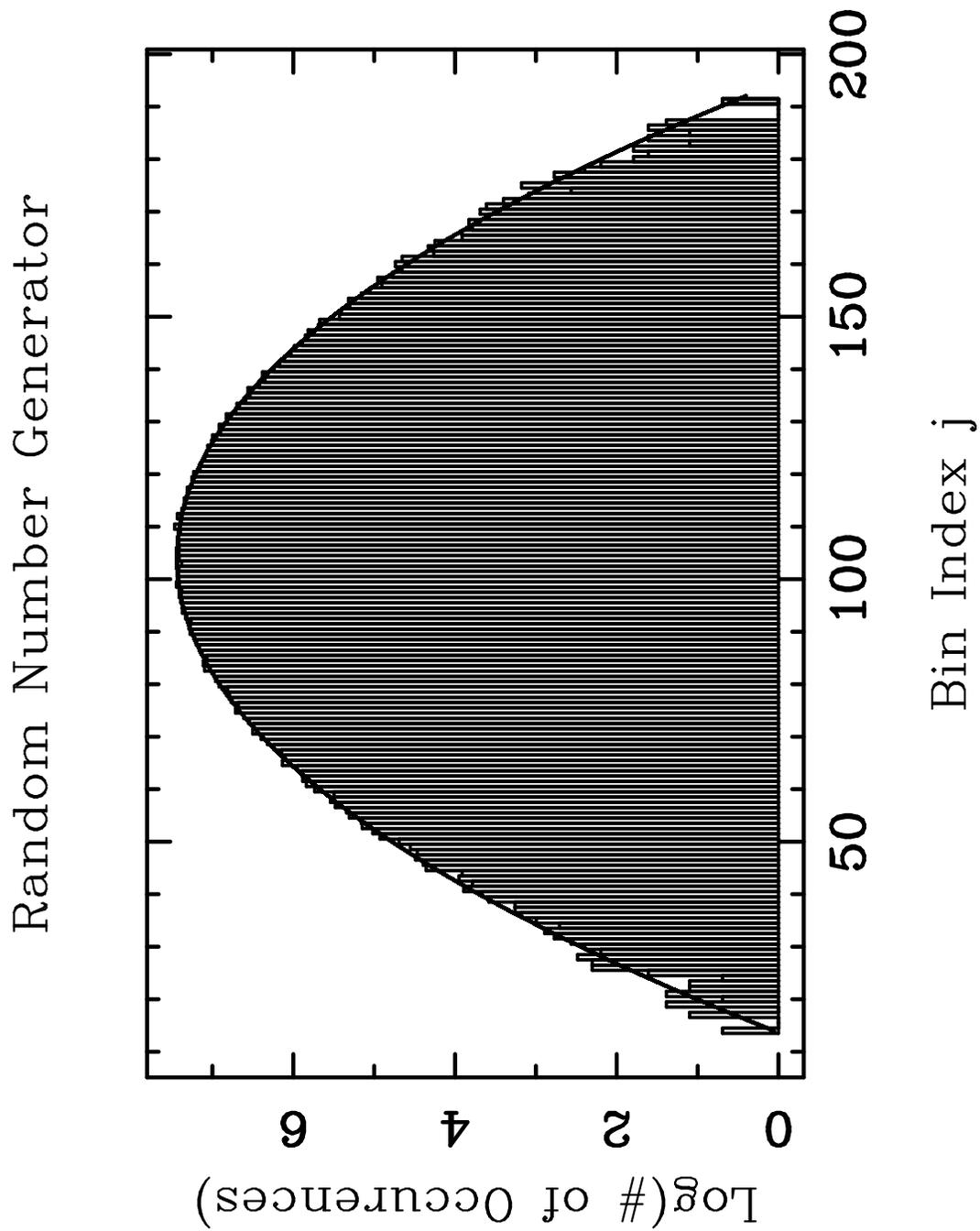}}
\vskip0.5in
\caption{Partition of the $10^5$ values obtained from the random 
number generator described in the text into 
200 bins. The logarithm of the number
of data points in each bin is plotted versus the bin number.
The solid parabola is a Gaussian fit.}
\label{fig:lewbar}
\end{figure}

\begin{figure}
\vskip0.5in
\centerline{\psfig{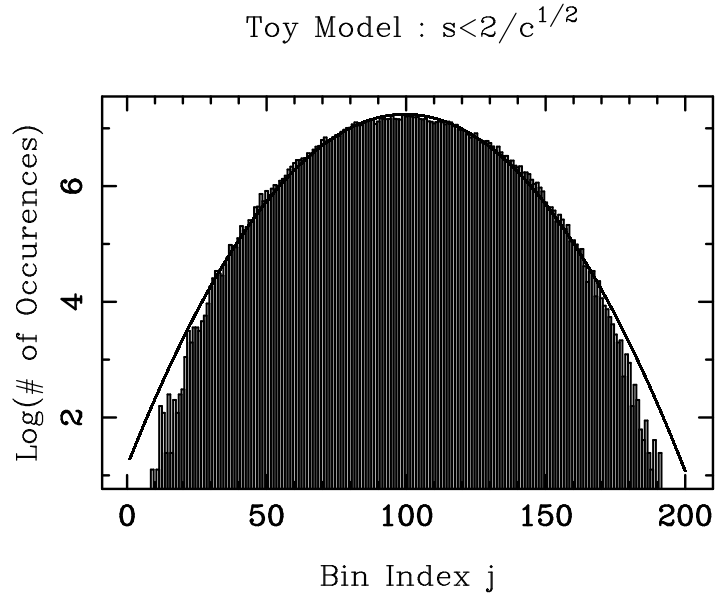}}
\vskip0.5in
\centerline{\psfig{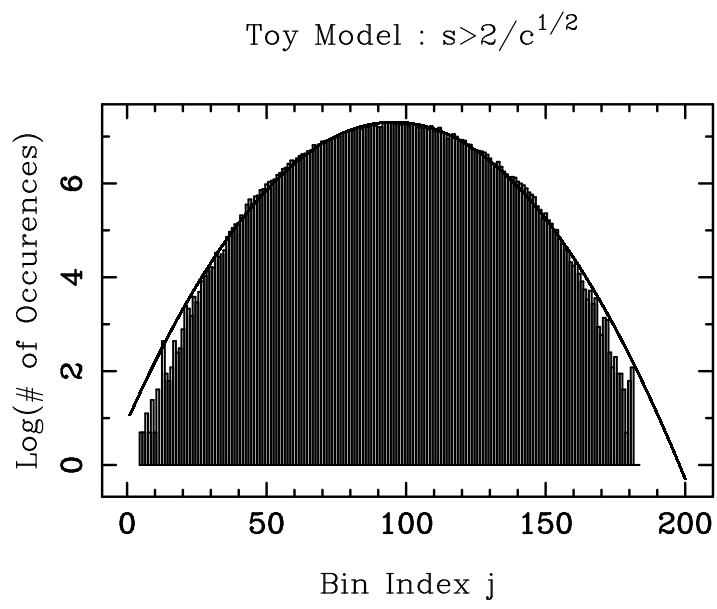}}
\vskip0.5in
\caption{Partition of the two sets of $10^5$ values described in 
subsection \ref{subsec:tmodel} into
200 bins, for the toy model. The logarithm of the number
of data points in each bin is plotted versus the bin number.
The solid parabola is a Gaussian fit.}
\label{fig:tbar}
\end{figure}

\begin{figure}
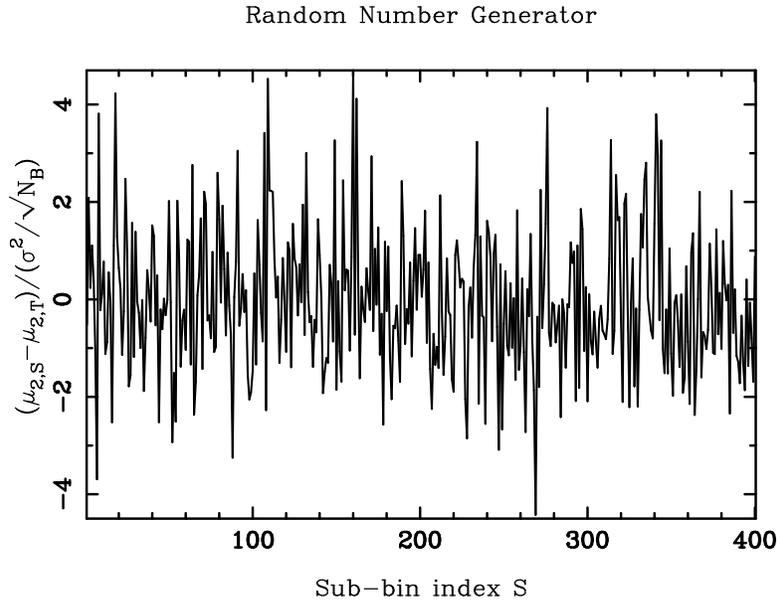

\vskip0.5in
\centerline{\psfig{figure=hier.ps,height=3.1in,angle=270}}
\vskip0.5in
\centerline{\psfig{figure=toy.ps,height=3.1in,angle=270}}
\hfill
\eject
\vskip0.5in
\centerline{\psfig{figure=lewis2.ps,height=3.1in,angle=270}}
\vskip0.5in
\caption{$(\widehat{\mu_{r,S}}-\widehat{\mu_{r,T}})/(\sigma^2/\sqrt{N_B})$
for 400 sub-bins indexed by $S$ as explained in the text, for 
the three sets previously abbreviated 
as H.M.a.+b., T.M.a.+b. and R.N.G.}
\label{fig:fluct}
\end{figure}
\vskip1.0in
\begin{table}
\label{tab:musi}
\caption{$\mu$ and $\sigma$ for the 7 sets of data considered}
\begin{tabular}{|c|ccccccc|}
 & H.M.(a.+b.) & H.M.a. & H.M.b. & T.M.(a.+b.) & T.M.a.&  T.M.b.& R.N.G.
\\ \hline
$\mu$    & 1045 & 1297 & 803 & -3037 & -3023 & -3029 & 4.22 \\  
$\sigma$ & 2123 & 2561 & 1524 & 3163.00 & 2062 & 3967& 4149
\end{tabular}
\end{table}
\vskip1.0in

\begin{table}
\caption{Values of the $p_i$ for the seven sets of data considered.}
\begin{tabular}{|c|ccccccc|} 
 & H.M.(a+b) & H.M.a. & H.M.b. & T.M.(a+b) & T.M.a.&  T.M.b. & R.N.G.
\\ \hline
p1 & 7.5 & 2.6 & 11 & 1.02 &0.86 & 0.85 & 1.057 \\  
p2 & 41 & 1.16 & 2.0 & 33 & 1.18 & 0.98 & 0.987 \\
p3 & 4.3 &  1.04 & 1.66 & 0.875 & 1.04 & 0.94 & 0.998 \\
p4 & 26 & 1.11 & 1.72 & 15 & 1.20 & 0.95 & 1.018
\end{tabular}
\end{table}
\vskip1.0in

\begin{table}
\caption{Skewness and kurtosis coefficients}
\begin{tabular}{|c|ccccccc|}
 & H.M.(a+b) & H.M.a. & H.M.b. & T.M.(a+b) & T.M.a.&  T.M.b.& R.N.G.
\\ \hline
$\mu_3/\mu_{2}^{3/2}$ & 0.31  & 0.12 & 0.15 & -0.0033 & -0.0090 & 0.0038
& -0.00048 \\
${\mu_4/\mu_2^2}-3$ & -0.58 & -1.10 & -0.81 & 0.56 & -0.29 & -0.34 
& -0.015  
\end{tabular}
\end{table}

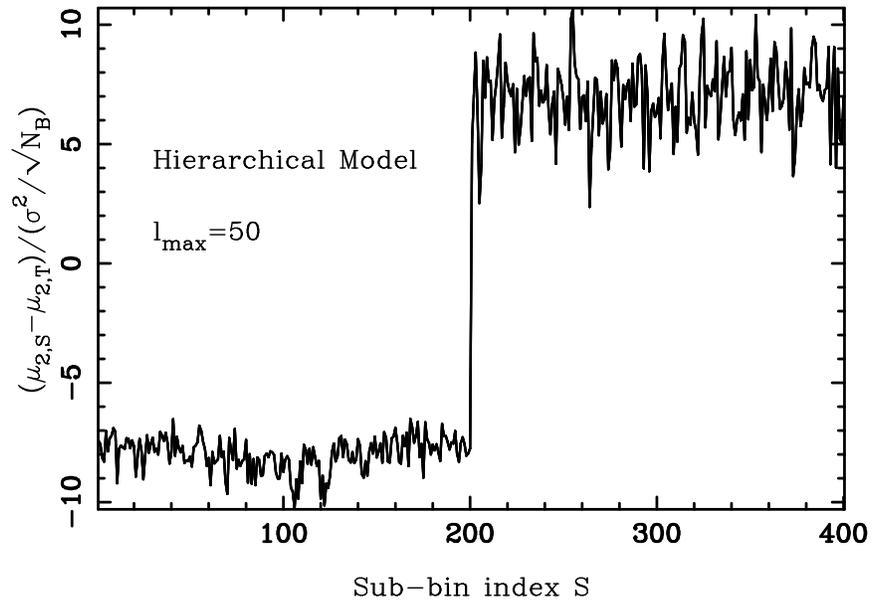
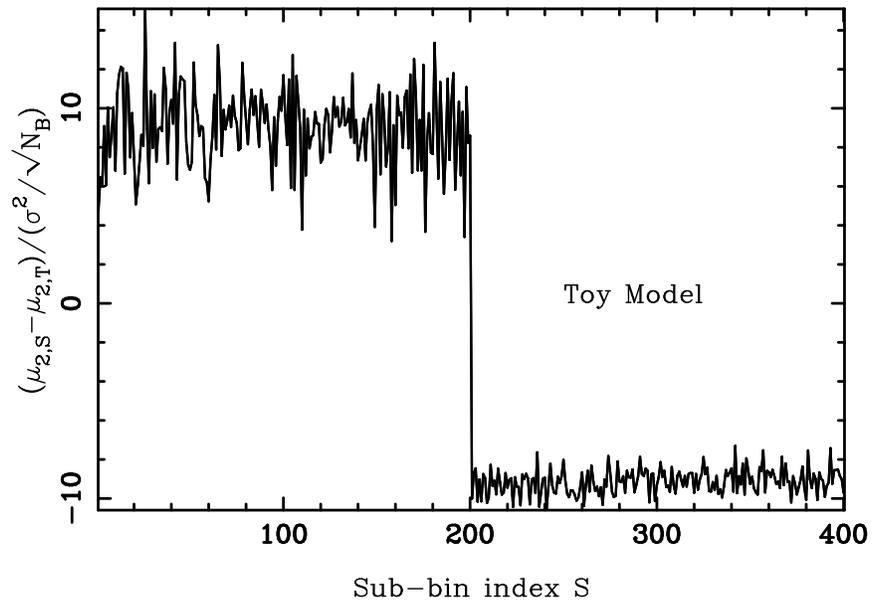
\end{document}